\documentclass[journal=jpccck,manuscript=article]{achemso}
\usepackage{graphicx}
\usepackage{dcolumn}
\usepackage{bm}
\usepackage[utf8]{inputenc}
\usepackage[T1]{fontenc}
\usepackage{mathptmx}
\usepackage{etoolbox}
\usepackage{gensymb}
\usepackage{booktabs}
\usepackage{amsmath} 
\usepackage{afterpage}
\usepackage{wrapfig}
\usepackage{float}
\usepackage{epstopdf}
\DeclareUnicodeCharacter{2212}{-}

\newcommand{\slfrac}[2]{\left.#1\middle/#2\right.}
\usepackage{wrapfig}
\usepackage{xcolor}
\usepackage[toc,page]{appendix}

\usepackage[version=3]{mhchem} 

\author{Abdullah El-Rifai}
\affiliation{Institute for Multiscale Thermofluids, University of Edinburgh, Edinburgh EH9 3FD, United Kingdom}
\email{abdullah.elrifai@hotmail.com}

\author{Liudmyla Klochko}%

\affiliation{ 
Universit\'e de Lorraine, CNRS, Inria, LORIA, 54000 Nancy, France
}

\author{Sreehari Perumanath}
\affiliation{Department of Mechanical Engineering, Indian Institute of Technology Madras, Chennai 600036, India}

\author{David Lacroix}%

\affiliation{ 
Universit\'e de Lorraine, CNRS, LEMTA, Nancy F-54000, France
}

\author{Rohit Pillai}
\email{r.pillai@ed.ac.uk}
\affiliation{Institute for Multiscale Thermofluids, University of Edinburgh, Edinburgh EH9 3FD, United Kingdom}

\author{Mykola Isaiev}%
 \email{mykola.isaiev@univ-lorraine.fr}
\affiliation{ 
Universit\'e de Lorraine, CNRS, LEMTA, Nancy F-54000, France
}

\title[]
  {Spectral Mechanisms of Solid/Liquid Interfacial Heat Transfer in the Presence of a Meniscus}

\abbreviations{SDHF}

\begin{document}





\begin{abstract}
In this study, we employ molecular simulations to investigate the enhancement in thermal conductance at the solid/liquid interface in the presence of a meniscus reported previously (Klochko et al., \textit{Phys. Chem. Chem. Phys.} 25(4):3298-3308, 2023). We vary the solid/liquid interaction strength at Lennard-Jones interfaces for both confined liquid and meniscus systems, finding that the presence of a meniscus yields an enhancement in the interfacial thermal conductance across all wettabilities. However, the magnitude of the enhancement is found to depend on the surface wettability, initially rising monotonously for low to moderate wettabilities, followed by a sharp rise between moderate and high wettabilities. The spectral decomposition of heat flux formalism was applied to understand the nature of this phenomenon further. By computing the in-plane and out-of-plane components of the heat fluxes within both the interfacial solid and liquid, we show that the initial monotonous rise in conductance enhancement predominantly stems from a rise in the coupling of out-of-plane vibrations within both the solid and the liquid. In contrast, the subsequent sharp rise at more wetting interfaces is linked to sharp increases in the utilization of the in-plane modes of the solid and liquid. These observations result from the interplay between the solid/liquid adhesive forces and the liquid/vapor interfacial tension. Our results can aid engineers in optimizing thermal transport at realistic interfaces, which is critical to designing effective cooling solutions for electronics, among other applications.

\end{abstract}

\section{Introduction}

The thermal transport across the region where the solid, liquid, and gas phases intersect, known as the ``three-phase contact line'' (TPCL), is key to phase-change processes,\cite{Raghupathi2016ContactInterface,Raghupathi2017PoolAugmentation} and impacts the performance of state-of-the-art two-phase cooling solutions for integrated circuits.\cite{Sefiane2022ProspectsChallenges} Recently, for example, candidate cooling devices have been developed where high heat flux dissipation is achieved by inducing evaporation from menisci pinned within nanoporous membranes. \cite{Hanks2018NanoporousManagement,Hanks2020HighMembranes} Given the importance of the heat transfer at the TPCL in such devices, as well as its relevance to other practical applications such as scanning thermal microscopy, \cite{Assy2014AnalysisMicroscopy,Bodzenta2020QuantitativeProbes,Zhang2020ASThM} phase change materials, \cite{Qiu2020Phase-ChangeApplications,Jilte2021AMaterials} and photo-thermally induced bubble growth, \cite{Yan2020AEngine} gaining a better understanding of the energy transport at such interfaces is becoming increasingly important.

Despite the nanoscale dimensions of the TPCL,\cite{carey2020liquid} several experimental and numerical studies have demonstrated that it plays a major role in evaporative heat transfer. \citet{Stephan1992AnalysisWalls} found that half of the heat transferred within a grooved heat pipe evaporator occurred across the TPCLs of the liquid menisci. \citet{Ibrahem2010ExperimentalLine} studied the heat transfer characteristics at the TPCL of a liquid meniscus pinned in a microchannel, finding that the peak evaporative heat flux occurs at the TPCL. \citet{Kunkelmann2012TheInvestigations} reported similar findings for advancing and receding menisci. \citet{Maroo2013FundamentalBoiling} simulated evaporation from nanoscale menisci, similarly reporting that the maximum evaporative heat flux occurs from the TPCL. Thus, to improve the rate of heat transfer in such systems, the energy transport at the TPCL must be optimized.

Numerous studies have reported that the overall heat transfer from evaporating TPCLs is impacted by the interfacial thermal conductance across the solid/liquid interface ($G$), which manifests as a temperature discontinuity at the interface ($\Delta T$), and is related to the cross-sectional area of the interface ($A$) and the heat transferred across it ($Q$) via $G = \slfrac{Q}{\Delta T}$. \citet{Zhao2011Near-wallMicrochannels} investigated evaporation processes in microchannels using kinetic theory, finding that $G$ can reduce the peak evaporative heat flux from the TPCL by approximately 15\%. \citet{Han2017Solid-LiquidTheory} and \citet{Ma2021MolecularSubstrate} studied thin-film evaporation using molecular dynamics (MD) simulations, finding that $G$ can bottleneck evaporative heat transfer by up to 20\% for regions of the TPCL where the liquid films become ultra-thin ($<100$ nm thick). Thus, a deeper understanding of $G$ in the presence of the TPCL can help optimize the rate of evaporative heat transfer in such processes.

\citet{Klochko2023MolecularMeniscus} investigated $G$ at the TPCL of a silicon/water interface for various surface wettabilities using a combination of MD and finite element method simulations. To understand how the presence of the TPCL influenced $G$, they quantified $G$ for two distinct systems: (i) a system in which the silicon walls fully confined the water and thus had an absence of vapor, referred to as the ``confined'' system, and (ii) a system that contained TPCLs induced by pinning a water meniscus between two silicon walls, denoted as the ``meniscus'' system. It is important to note that, in this configuration, the liquid in the meniscus system had a reduced area of contact with the solid. Thus, to compare $G$ across the two systems, \citet{Klochko2023MolecularMeniscus} normalized $G$ by the solid/liquid contact area $A_{\rm{c}}$. They found that the presence of a meniscus led to an enhancement in normalized $G$ across all surface wettabilities, implying that solid/liquid energy transport is facilitated in the presence of the TPCL. However, the origins of this enhancement in interfacial heat transfer have not yet been understood.

While studies on the enhancement of $G$ at the TPCL are sparse, efforts have been made to uncover the origins of enhancements in $G$ at solid/liquid interfaces through the analysis of the \textit{structural properties} of the interfacial liquid. \citet{Shenogina2009HowInterfaces} observed that $G$ exhibits a linear relationship with the work of adhesion at self-assembled-monolayer/water interfaces; however, other studies have reported that this cannot be used as the sole predictor of $G$. \cite{Ge2013VibrationalInterface,Tomko2019NanoscaleInterfaces,Gonzalez-Valle2021MolecularInterfaces}. \citet{Alexeev2015KapitzaEffects} reported a correlation between the peak interfacial density of the liquid and $G$ at a graphene/water interface, but other studies have shown that this correlation does not always hold.\cite{Ma2018OrderedWater,Peng2022ReducingStructure,Li2020AtomicSimulation,Alosious2022EffectsInterfaces,Han2017ThermalStructuring,Li2022RoleInterface,Gonzalez-Valle2021MolecularInterfaces} \citet{Ramos-Alvarado2016Solid-LiquidStructure} demonstrated a correlation between the density depletion length and $G$ at a silicon/water interface, however this could not be successfully applied to other interfaces.\cite{Ramos-Alvarado2017SpectralInterfaces,Gonzalez-Valle2018ThermalInterfaces,Gonzalez-Valle2019SpectralInterfaces,Li2020AtomicSimulation,Li2022RoleInterface} \citet{Ma2018OrderedWater} demonstrated that the increase in $G$ due to the presence of surface charges at a graphene/water interface was linked to a rise in the in-plane ordering of the interfacial liquid, but other studies have shown that this is not universally applicable.\cite{Li2020AtomicSimulation,Peng2022ReducingStructure,Anandakrishnan2023EffectsInterface} While structural properties have been partially successful in explaining increases in $G$, their lack of universality means other approaches must be sought to fully understand the origins of enhancements in $G$.

An alternative method to gain insights into enhancements in $G$ at solid/liquid interfaces revolves around deploying \textit{spectral techniques} to directly probe the behavior of the interfacial atomic vibrations, the sole energy carriers across electrically non-conductive interfaces.\cite{Chen2022InterfacialFuture} In a number of studies, the enhancements in $G$ were attributed to a change in the \textit{frequencies} of the vibrations engaged in interfacial heat transfer, referred to as ``utilized'' modes. \citet{Qian2018LowerSurface} found that the rise in $G$ with increasing surface wettability at a graphene/ionic-liquid interface was linked to a rise in the utilization of high-frequency modes within the interfacial solid. \citet{Ma2018OrderedWater} demonstrated that the enhancement in $G$ yielded by an increase in the magnitude of surface charges at a graphene/water interface is similarly associated with greater high-frequency mode utilization in the interfacial solid. \citet{Qian2019UltralowLayer} reported a comparable effect at graphene/ionic-liquid interfaces with increasing surface charge magnitude. In our previous work,\cite{El-Rifai2024UnravelingInterface} we demonstrated that the exponential-to-linear regime cross-over that $G$ experiences with increasing solid/liquid interaction strength ($\epsilon_{\rm{SL}}$) at LJ interfaces is related to an increase in the similarity of the frequencies of the utilized vibrations within the interfacial solid and liquid. The successful implementation of spectral techniques to understand enhancements in $G$ at solid/liquid interfaces in these studies raises the following question for the TPCL system of interest here: \textit{does the presence of a TPCL similarly alter the frequencies of the vibrations utilized, and does that spectral variation enhance normalized $G$?}

In other studies, the increases in $G$ at solid/liquid interfaces were also found to be linked to the \textit{orientation} of the utilized vibrations with respect to the plane of the interface. Note that vibrational modes can be classified by orientation into: (i) in-plane (i.e. parallel to the interface), and (ii) out-of-plane (i.e. perpendicular to the interface) vibrations. \citet{Ramos-Alvarado2017SpectralInterfaces} found that the rise in $G$ with increasing wettability at silicon/water interfaces was linked to greater high-frequency mode utilization, as well as increased in-plane mode utilization in the solid. \citet{Gonzalez-Valle2019SpectralInterfaces} reported similar findings at a silicon-carbide/water interface. \citet{Zhou2021EffectMonolayer} demonstrated that the rise in $G$ induced by the presence of atomic defects at a graphene/hydrocarbon interface is similarly associated with greater in-plane mode utilization in the solid. These studies raise the second question about the TPCL-induced enhancement in the area-normalized $G$: \textit{does the presence of a TPCL influence the orientation of the vibrations utilized, and does that spectral variation enhance normalized $G$?}

\citet{Klochko2023MolecularMeniscus} hypothesized that the enhancement in normalized $G$ in the presence of the TPCL stems from greater in-plane mode utilization in the interfacial solid; however, this hypothesis was not quantified using spectral techniques. In our previous work\cite{El-Rifai2024UnravelingInterface}, the exponential-to-linear regime cross-over in $G$ was explained by analyzing both the modes utilized within the solid to \textit{transmit} energy to the liquid, as well as the modes within the liquid used to \textit{receive} energy from the solid at a Lennard-Jones (LJ) interface. Therefore, in this work, the vibrations utilized within both the solid and liquid will be analyzed for the first time to understand the origin of the enhancement in normalized $G$ yielded by the presence of a TPCL at a simple LJ interface.

\section{Methodology}

\subsection{System setup}

\begin{figure}[h!]
\centering
\includegraphics[width=0.5\linewidth]{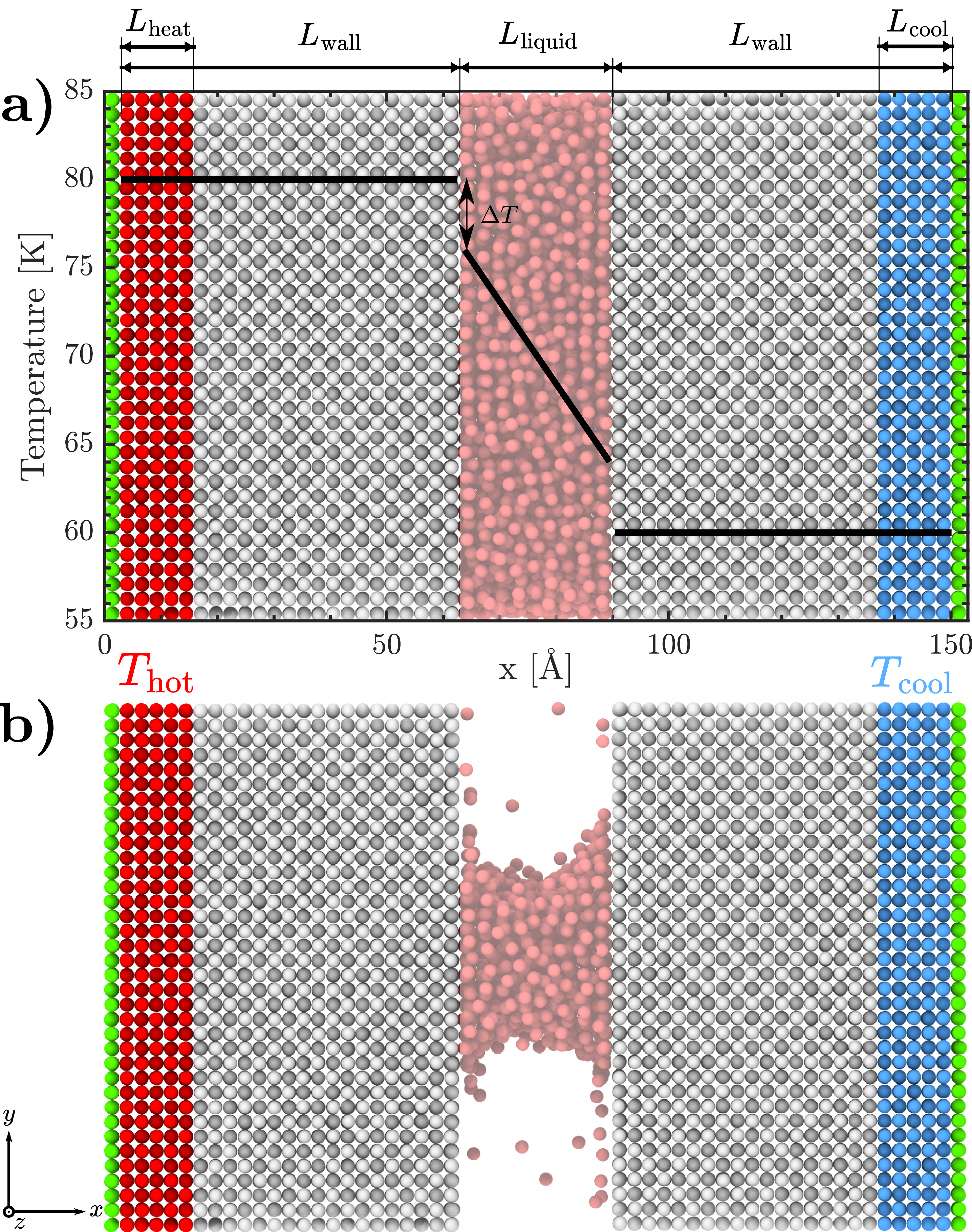}
\caption{\label{fig:PROJ2setup}(a) A nanochannel comprising two LJ FCC [100] walls of length $L_{\rm{wall}}$ separated by a distance $L_{\rm{liquid}}$, completely filled with a Lennard-Jones (LJ) liquid. The temperature gradient within all materials is shown using solid black lines, with the solid/liquid interfacial discontinuity $\Delta T$ highlighted using black arrows. (b) The same nanochannel is only partially filled with an LJ meniscus. In both cases, the hot and cold walls are thermostatted to $T_{\rm{hot}}=80$K and $T_{\rm{cold}}=60$K respectively.}
\end{figure}

We conduct non-equilibrium MD simulations using LAMMPS \cite{Thompson2022LAMMPSScales} to investigate the origin of the enhancement in interfacial energy transport at an LJ interface. We consider two systems: (i) a liquid fully confined between the walls of a nanochannel, referred to as the ``confined'' system (Figure \ref{fig:PROJ2setup}(a)), and (ii) a partially filled nanochannel comprising a meniscus, referred to as the ``meniscus'' system (Figure \ref{fig:PROJ2setup}(b)). The system geometry is identical for both the confined and meniscus systems. The nanochannel width is set to $L_{\rm{liquid}} \approx 30$ \AA. The cross-sectional area of the domain $A_{\rm{wall}}$ is set to $18 \times 18$ $a^{2}$, where $a=5.34$ \AA\ is the lattice constant of the solid. The wall thickness is set to $L_{\rm{wall}}=12a$.

Simulations are initiated with an equilibration stage at 70K during 5 ns for both systems. In the case of the confined system, the pressure is controlled by applying a piston to the right wall to achieve the pressure corresponding to 1 atm. After the equilibration stage, the piston condition is removed, and the outermost green-colored solid layers at both ends of the simulation domain (see Figure \ref{fig:PROJ2setup}) are held rigid to maintain the specified pressure as well as avoid the displacement of the interfaces. The red-colored solid atoms in the left wall (tagged as $L_{\rm{heat}}$ in Figure \ref{fig:PROJ2setup}) are then set to $T_{\rm{hot}}=80$K, while the blue-colored atoms in the right wall (tagged as $L_{\rm{cool}}$ in Figure \ref{fig:PROJ2setup}) are set to $T_{\rm{cold}}=60$K, using a Nos\'e-Hoover thermostat. The gray portions of both walls are permitted to vibrate freely using the NVE ensemble.

\begin{figure}[htb!]
\centering
\includegraphics[width=0.5\linewidth]{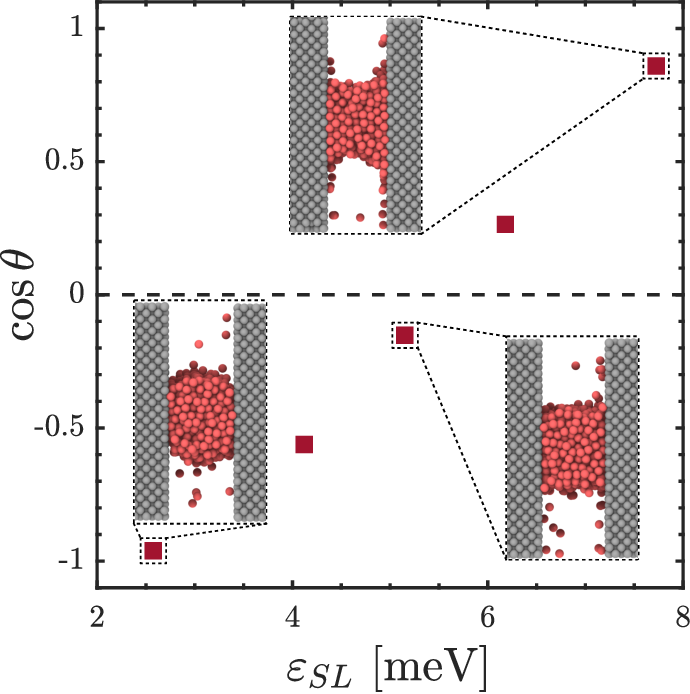}
\caption{The relationship between the cosine of the contact angle that the liquid makes with the surface ($\theta$) and $\epsilon_{\rm{SL}}$. The insets depict the shape of the meniscus at various values of $\epsilon_{\rm{SL}}$.}
\label{fig:contactangles}
\end{figure}

All interatomic interactions are modeled using the LJ potential\cite{allen2017computer} using a cut-off distance $r_{\rm{cut}} = 8.5$ \AA. In line with prior studies on LJ interfaces,\cite{Saaskilahti2016SpectralInterfaces,Giri2016ImplicationsStudy,El-Rifai2024UnravelingInterface} the liquid/liquid interaction strength is set to $\epsilon_{\rm{LL}}=10.3$ meV, while the solid/solid interaction strength is set to $\epsilon_{\rm{SS}}=\epsilon_{\rm{LL}}$. To quantify the impact of the surface wettability on $\Gamma$, the solid/liquid interaction strength $\epsilon_{\rm{SL}}$ is varied directly in the range from 2.5 meV to 7.5 meV. The resulting wetting angles ($\theta$) corresponding to all values of $\epsilon_{\rm{SL}}$ were calculated from the density profiles similar to \citet{isaiev2016}, and are presented in Figure \ref{fig:contactangles}. MD snapshots of the resulting meniscus shape are illustrated in the insets. The range of values for $\epsilon_{\rm{SL}}$ is chosen such that menisci of various curvatures are produced, with the transition from phobicity to philicity occurring around $\epsilon_{\rm{SL}} \approx 5$ meV, as observed in Figure \ref{fig:contactangles}. The maximum value of $\epsilon_{\rm{SL}}$ is chosen so the meniscus does not thoroughly wet the surface.

\subsection{Calculation of the Interfacial Thermal Conductance}

The presence of thermostats at both ends of the domain which are 20K apart produces a linearly varying temperature distribution within the domain (shown using black lines in Figure \ref{fig:PROJ2setup}(a)). The consequent interfacial temperature discontinuity at the solid/liquid interface ($\Delta T$) can be measured by extrapolating the best-fit line through the bulk liquid's temperature field to the interfacial solid. The normalized interfacial thermal conductance is then computed using

\begin{equation}
\label{eq:gammapr}
\Gamma = \frac{Q}{\Delta T A_{\rm{c}}},  
\end{equation}

\noindent where $Q$ is the heat transferred across the interface, and $A_{\rm{c}}$ is the actual solid/liquid contact area. Note that in the confined case, $A_{\rm{c}}$ is equivalent to the cross-sectional area of the wall $A_{\rm{wall}}$, i.e. $A_{\rm{c}} = A_{\rm{wall}}$, while in the meniscus case $A_{\rm{c}} < A_{\rm{wall}}$.

\subsection{Spectral Analysis of the Interfacial Thermal Conductance}

\begin{figure}[htb!]
\centering
\includegraphics[width=0.35\linewidth]{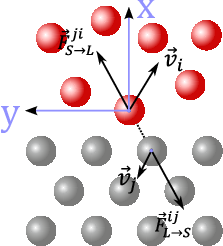}
\caption{Schematic representation of the interactions between atoms of a solid and a liquid.}
\label{fig:forceschematic}
\end{figure}

The spectral analysis of interfacial heat transfer can be conducted using the methodology developed by \citet{Saaskilahti2016SpectralInterfaces}, and validated for liquids of varying wettability in our previous work. \cite{El-Rifai2024UnravelingInterface} Through this, the spectral decomposition of heat flux within the interfacial solid $q(\omega)_{\rm{L \rightarrow S}}$ can be computed, which quantifies the contributions of each vibration of frequency $\omega$ utilized within the solid to \textit{transmit} energy to the liquid. It can be derived from the Fourier transform of the cross-correlation between the cumulative forces that all liquid atoms exert upon each solid atom and the velocities of the solid atoms via: 

\begin{equation}
    q(\omega)_{\rm{L \rightarrow S}} = \frac{2}{A_c} \cdot \Re \left( \sum_{i \in L}\sum_{j \in S} \int_{-\infty}^{\infty} d \tau \rm{e}^{i \omega \tau} \left< \overrightarrow{F}_{\rm{L \rightarrow S}}^{ij}(\tau) \cdot \overrightarrow{v}_j(0) \right> \right),
\end{equation}

\noindent where $\tau$ is the correlation time between forces and velocities; $\overrightarrow{F}_{\rm{L \rightarrow S}}^{ij}$ is the force acting on the $j$-th atom of the solid from the $i$-th atom of the liquid; and $\overrightarrow{v}_i$ and $\overrightarrow{v}_j$ are the velocities of the $i$-th and $j$-th atom respectively, as illustrated in Figure \ref{fig:forceschematic}.

The spectral decomposition of heat flux within the interfacial liquid $q(\omega)_{\rm{S \rightarrow L}}$, which instead quantifies the contributions of each vibration of frequency $\omega$ utilized within the liquid to \textit{receive} energy from the solid, can be calculated using a similar approach. It can be derived from the Fourier transform of the cross-correlation between the cumulative forces that all solid atoms exert upon each liquid atom and the velocities of the liquid atoms via:  

\begin{equation}
    q(\omega)_{\rm{S \rightarrow L}} = \frac{2}{A_c} \cdot \Re \left( \sum_{i \in L}\sum_{j \in S} \int_{-\infty}^{\infty} d \tau \rm{e}^{i \omega \tau} \left< \overrightarrow{F}_{\rm{S \rightarrow L}}^{ji}(\tau) \cdot \overrightarrow{v}_i(0) \right> \right),
\end{equation}

\noindent where $\tau$ is the correlation time between forces and velocities; $\overrightarrow{F}_{\rm{S \rightarrow L}}^{ji}$ is the force acting on the $i$-th atom of the liquid from the $j$-th atom of the solid; and $\overrightarrow{v}_i$ and $\overrightarrow{v}_j$ are the velocities of the $i$-th and $j$-th atom respectively, similarly depicted in Figure \ref{fig:forceschematic}.

In order to consider the \textit{orientation} of the vibrations, rather than merely their \textit{frequency}, the directional components of the force and velocity vectors must be considered independently. The utilization of out-of-plane vibrations ($q_{\rm{L \rightarrow S}}^{\perp}$ and $q_{\rm{S \rightarrow L}}^{\perp}$) is computed using the component of the forces and velocities vectors that is perpendicular to the interface, via: 

\begin{equation}
    q(\omega)_{\rm{L \rightarrow S}}^{\perp} = \frac{2}{A_c} \cdot \Re \left( \sum_{i \in L}\sum_{j \in S} \int_{-\infty}^{\infty} d \tau \rm{e}^{i \omega \tau} \left< \left( \overrightarrow{F}_{\rm{L \rightarrow S}}^{ij}(\tau) \right)_x \cdot \left( \overrightarrow{v}_j(0) \right)_x \right> \right),
\end{equation}

\noindent and

\begin{equation}
    q(\omega)_{\rm{S \rightarrow L}}^{\perp} = \frac{2}{A_c} \cdot \Re \left( \sum_{i \in L}\sum_{j \in S} \int_{-\infty}^{\infty} d \tau \rm{e}^{i \omega \tau} \left< \left( \overrightarrow{F}_{\rm{S \rightarrow L}}^{ji}(\tau) \right)_x \cdot \left( \overrightarrow{v}_i(0) \right)_x \right> \right).
\end{equation}

Similarly, the utilization of in-plane modes ($q_{\rm{L \rightarrow S}}^{\parallel}$ and $q_{\rm{S \rightarrow L}}^{\parallel}$) is obtained from the components of the force and velocity vectors that are parallel to the interface, as follows:

\begin{equation}
    \label{eq:qLtoS}
    \begin{split}
        q(\omega)_{\rm{L \rightarrow S}}^{\parallel} = \frac{2}{A_c} \cdot \Re \Bigg( \sum_{i \in L}\sum_{j \in S} \int_{-\infty}^{\infty} d \tau \rm{e}^{i \omega \tau} \Big( \left< \left( \overrightarrow{F}_{\rm{L \rightarrow S}}^{ij}(\tau) \right)_y \cdot \left( \overrightarrow{v}_j(0) \right)_y \right> \\
        + \left< \left( \overrightarrow{F}_{\rm{L \rightarrow S}}^{ij}(\tau) \right)_z \cdot \left( \overrightarrow{v}_j(0) \right)_z \right> \Big) \Bigg),
    \end{split}
\end{equation}
\noindent and

\begin{equation}
    \label{eq:qStoL}
    \begin{split}
        q(\omega)_{\rm{S \rightarrow L}}^{\parallel} = \frac{2}{A_c} \cdot \Re \Bigg( \sum_{i \in L}\sum_{j \in S} \int_{-\infty}^{\infty} d \tau \rm{e}^{i \omega \tau} \Big( \left< \left( \overrightarrow{F}_{\rm{S \rightarrow L}}^{ji}(\tau) \right)_y \cdot \left( \overrightarrow{v}_i(0) \right)_y \right> \\
        + \left< \left( \overrightarrow{F}_{\rm{S \rightarrow L}}^{ji}(\tau) \right)_z \cdot \left( \overrightarrow{v}_i(0) \right)_z \right> \Big) \Bigg).
    \end{split}
\end{equation}

From the separate components of the $q(\omega)$, the cumulative thermal conductance across the solid/liquid interface can be calculated for both in-plane and out-of-plane modes as follows:

\begin{equation}
\label{eq:decomp}
\Gamma (\omega)_{\rm{L \rightarrow S, S \rightarrow L}}^{\perp, \parallel}  = \frac{1}{\Delta T} \int_{0}^{\omega}q(\omega)_{\rm{L \rightarrow S, S \rightarrow L}}^{\perp, \parallel}  \frac{d \omega}{2 \pi},
\end{equation}

\noindent and the total one:

\begin{equation}
\Gamma (\omega)_{\rm{L \rightarrow S, S \rightarrow L}}  = \Gamma (\omega)_{\rm{L \rightarrow S, S \rightarrow L}}^{\perp}  + \Gamma (\omega)_{\rm{L \rightarrow S, S \rightarrow L}}^{\parallel} .
\end{equation}

\section{Results \& Discussion}

\subsection{Effect of Meniscus on $\Gamma$ at Various Values of $\epsilon_{\rm{SL}}$}

\begin{figure}[htb!]
\centering
\includegraphics[width=\linewidth]{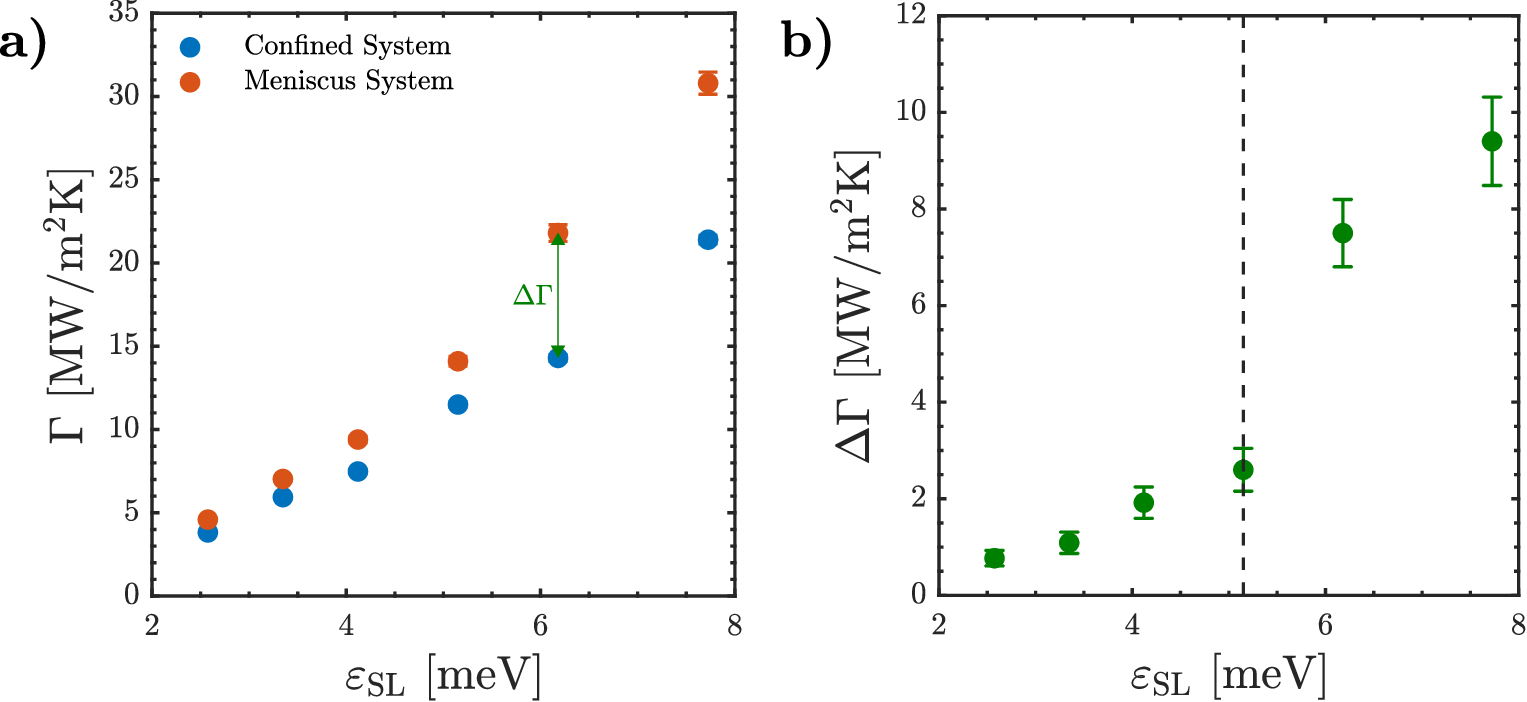}
\caption{a) The interfacial thermal conductance normalized by the actual liquid contact area ($\Gamma$) for the confined (blue points) and meniscus (orange points) systems at various values of the solid/liquid interaction strength $\epsilon_{\rm{SL}}$. b) The increase in $\Gamma$ due to the presence of a meniscus ($\Delta\Gamma$) for different values of $\epsilon_{\rm{SL}}$.}
\label{fig:PROJ2Gammameniscusconfined}
\end{figure}

In Figure \ref{fig:PROJ2Gammameniscusconfined}(a), the variation of $\Gamma$ with $\epsilon_{\rm{SL}}$ is presented for the meniscus and confined systems as obtained through Equation \ref{eq:gammapr}. With rising $\epsilon_{\rm{SL}}$, $\Gamma$ increases steadily for both systems. However, it can be observed that the magnitude of $\Gamma$ in the meniscus system is larger than that of the confined system. This implies that the presence of a meniscus yields an enhancement in $\Gamma$ across all values of $\epsilon_{\rm{SL}}$. This enhancement is denoted as $\Delta \Gamma =\Gamma_{\rm{meniscus}} - \Gamma_{\rm{confined}}$, and is illustrated for the $\epsilon_{\rm{SL}} \approx 6$ meV case in Figure \ref{fig:PROJ2Gammameniscusconfined}(a). Thus, we have successfully reproduced the enhancement in normalized $G$ reported at a silicon/water interface by \citet{Klochko2023MolecularMeniscus} using a simple LJ interface for the first time, strengthening the case for using our setup to understand the spectral origins of the enhancement in $\Gamma$.

Next, $\Delta \Gamma$ is calculated for all values of $\epsilon_{\rm{SL}}$, as shown in Figure \ref{fig:PROJ2Gammameniscusconfined}(b). With increasing $\epsilon_{\rm{SL}}$, $\Delta \Gamma$ is seen to increase monotonously until $\epsilon_{\rm{SL}} \approx 5$ meV. However, there is a discontinuity at $\epsilon_{\rm{SL}} \approx 5$ meV, and it can be observed that $\Delta \Gamma$ experiences a sharp increase beyond this point. This indicates that, while the presence of a meniscus enhances $\Gamma$ regardless of $\epsilon_{\rm{SL}}$, the magnitude of this enhancement is dependent on $\epsilon_{\rm{SL}}$. From Figure \ref{fig:contactangles}, we know that the  transition from phobicity to philicity occurs around $\epsilon_{\rm{SL}} \approx 5$ meV. Therefore, the sharp rise in $\Delta \Gamma$ observed at this point coincides with a change in the characteristic curvature of the liquid/vapor interface, namely from concavity to convexity in the direction of the vapor. This raises the third question to be addressed in this work: \textit{why does the enhancement in $\Gamma$ experience a sharp enhancement beyond $\epsilon_{\rm{SL}} \approx 5$ meV?}

To answer the questions posed thus far, we will first assess whether the presence of a TPCL alters the \textit{frequencies} of the utilized modes within both the interfacial solid and liquid, followed by an examination of the \textit{orientation} of the utilized vibrations. This will be achieved by analyzing the spectral mechanisms of interfacial energy transport within the meniscus system, and comparing them to those within the benchmark confined system at each wettability.

\subsection{Effect of Meniscus on Frequencies of Utilized Vibrations}

In Figure \ref{fig:appendixnonnormalised} of the Appendix, the $q(\omega)_{\rm{L \rightarrow S}}$ and $q(\omega)_{\rm{S \rightarrow L}}$ distributions are presented for the confined and meniscus systems at three values of $\epsilon_{\rm{SL}}$. It can be observed that the the magnitudes of the $q(\omega)_{\rm{L \rightarrow S}}$ and $q(\omega)_{\rm{S \rightarrow L}}$ distributions differ greatly across the two systems, because the meniscus and confined systems possess significantly different magnitudes of $\Gamma$ for the same value of $\epsilon_{\rm{SL}}$. Attempting to compare two distributions of vastly different magnitudes would obscure any differences between them. Therefore, the $q(\omega)_{\rm{L \rightarrow S}}$ and $q(\omega)_{\rm{S \rightarrow L}}$ distributions of the two systems are normalized by setting the area beneath their curves to unity to facilitate their comparison in the remainder of this section. First, the normalized $q(\omega)_{\rm{L \rightarrow S}}$ distributions of the meniscus and confined systems will be compared, followed by their normalized $q(\omega)_{\rm{S \rightarrow L}}$ distributions.

\subsubsection{Comparing the $q(\omega)_{\rm{L \rightarrow S}}$ Distributions}

\begin{figure}[htb!]
\centering
\includegraphics[width=\linewidth]{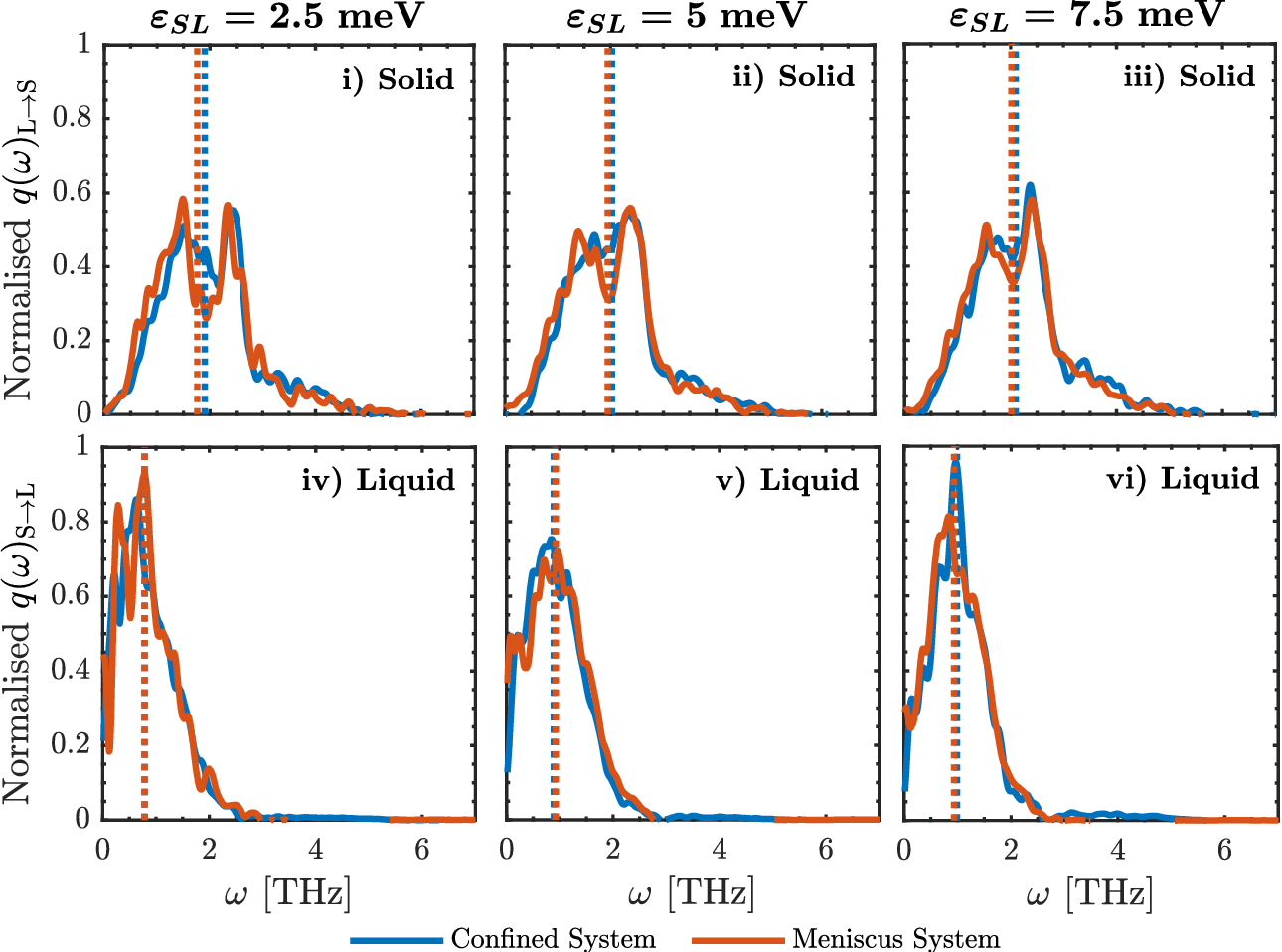}
\caption{Normalized spectral decompositions of heat flux within the interfacial solid ($q(\omega)_{\rm{L \rightarrow S}}$) and liquid ($q(\omega)_{\rm{S \rightarrow L}}$) for the confined and meniscus systems at select values of the solid/liquid interaction strength $\epsilon_{\rm{SL}}$.}
\label{fig:PROJ2SDHFnormalised}
\end{figure}

The normalized distributions of the spectral decompositions of heat flux within the interfacial solid ($q(\omega)_{\rm{L \rightarrow S}}$) for both the confined (blue line) and meniscus (orange line) systems are shown in Figures \ref{fig:PROJ2SDHFnormalised}(i)-(iii). Three values of $\epsilon_{\rm{SL}}$ from the cases presented in Figure \ref{fig:PROJ2Gammameniscusconfined} are selected, with the dotted blue and and orange lines representing the median value of that distribution. From Figures \ref{fig:PROJ2SDHFnormalised}(i)-(iii), it can be observed that there are no statistically significant differences in the distributions of $q(\omega)_{\rm{L \rightarrow S}}$ between the meniscus and confined systems.

\subsubsection{Comparing the $q(\omega)_{\rm{S \rightarrow L}}$ Distributions}

Figures \ref{fig:PROJ2SDHFnormalised}(iv)-(vi) illustrate the normalized spectral decomposition of heat flux within the interfacial liquid ($q(\omega)_{\rm{S \rightarrow L}}$) for both the confined and meniscus systems at the same magnitudes of $\epsilon_{\rm{SL}}$. From these plots, it is apparent that the normalized distributions of $q(\omega)_{\rm{S \rightarrow L}}$ are virtually identical for both systems across all values of $\epsilon_{\rm{SL}}$ as demonstrated by their near-perfect overlap, similar to the case of the interfacial solid. Once again, this means that the frequencies of the modes utilized within the interfacial liquid are similar across the two systems, regardless of $\epsilon_{\rm{SL}}$. Consequently, the enhancement in $\Gamma$ due to the presence of a meniscus cannot be explained by a change in the frequencies of the vibrations utilized within the interfacial solid or liquid, and alternative mechanisms must be explored.

\subsection{Effect of Meniscus on Orientation of Utilized Vibrations}

Rather than merely comparing the overall spectral distributions of $q(\omega)_{\rm{L \rightarrow S}}$ and $q(\omega)_{\rm{S \rightarrow L}}$, which probes whether the meniscus and confined systems utilize modes of different \textit{frequencies}, these spectra can be further decomposed based on the \textit{orientation} of the vibrations with respect to the plane of the interface.\cite{Saaskilahti2016SpectralInterfaces} As stated previously, differences in the orientation of utilized modes have been observed in the literature,\cite{Ramos-Alvarado2017SpectralInterfaces,Gonzalez-Valle2019SpectralInterfaces,Zhou2021EffectMonolayer} but this has not yet been studied at the TPCL. Thus, the next step is to analyze the contribution of the in-plane and out-of-plane modes towards the interfacial thermal conductance $\Gamma$.

The  spectral distributions $q(\omega)_{\rm{L \rightarrow S}}$ and $q(\omega)_{\rm{S \rightarrow L}}$ can be decomposed into their out-of-plane ($q(\omega)_{\rm{L \rightarrow S}}^{\perp}$,$q(\omega)_{\rm{S \rightarrow L}}^{\perp}$) and in-plane ($q(\omega)_{\rm{L \rightarrow S}}^{\parallel}$,$q(\omega)_{\rm{S \rightarrow L}}^{\parallel}$) components using Equations \ref{eq:qLtoS}-\ref{eq:qStoL}, as presented in Figures \ref{fig:appendixperp}-\ref{fig:appendixparallel} of the Appendix. However, this merely quantifies the contributions of each \textit{individual} vibrational frequency $\omega$ towards interfacial energy transport. An alternative metric is the \textit{cumulative} contribution of the entire \textit{range} of vibrational frequencies towards $\Gamma$. This can be obtained by computing the ``spectral accumulation'' of each distribution, calculated via their cumulative integration with respect to $\omega$, as presented in Equation \ref{eq:decomp}. Using this measurement, the separate contributions of in-plane and out-of-plane modes towards the total energy transported in each system can be compared across the meniscus and confined systems. Note that, as we are now interested in comparing the actual energetic contributions of each vibrational orientation rather than merely their spectral distribution, the spectral accumulations will not be normalized. First, this analysis will be conducted in the interfacial solid, followed by the interfacial liquid.

The representative examples of the cumulative spectral distribution for the confined liquid and the meniscus case are presented in Fig.\ref{fig:PROJ2SDHFnonnormalised}(a) and Fig.\ref{fig:liquidaccum}(a). It should be noted that both cumulative spectral distributions under higher frequencies are equal to the thermal conductance calculated by the use of Equation \ref{eq:gammapr}: $ \Gamma = \Gamma(\omega \rightarrow \infty)_{\rm{L \rightarrow S}} = \Gamma(\omega \rightarrow \infty)_{\rm{S \rightarrow L}}$ (see Figure \ref{fig:appendixvalidation} in the Appendix).

\subsubsection{Spectral Accumulations within the Interfacial Solid}

\begin{figure}[t!]
\centering
\includegraphics[width=\linewidth]{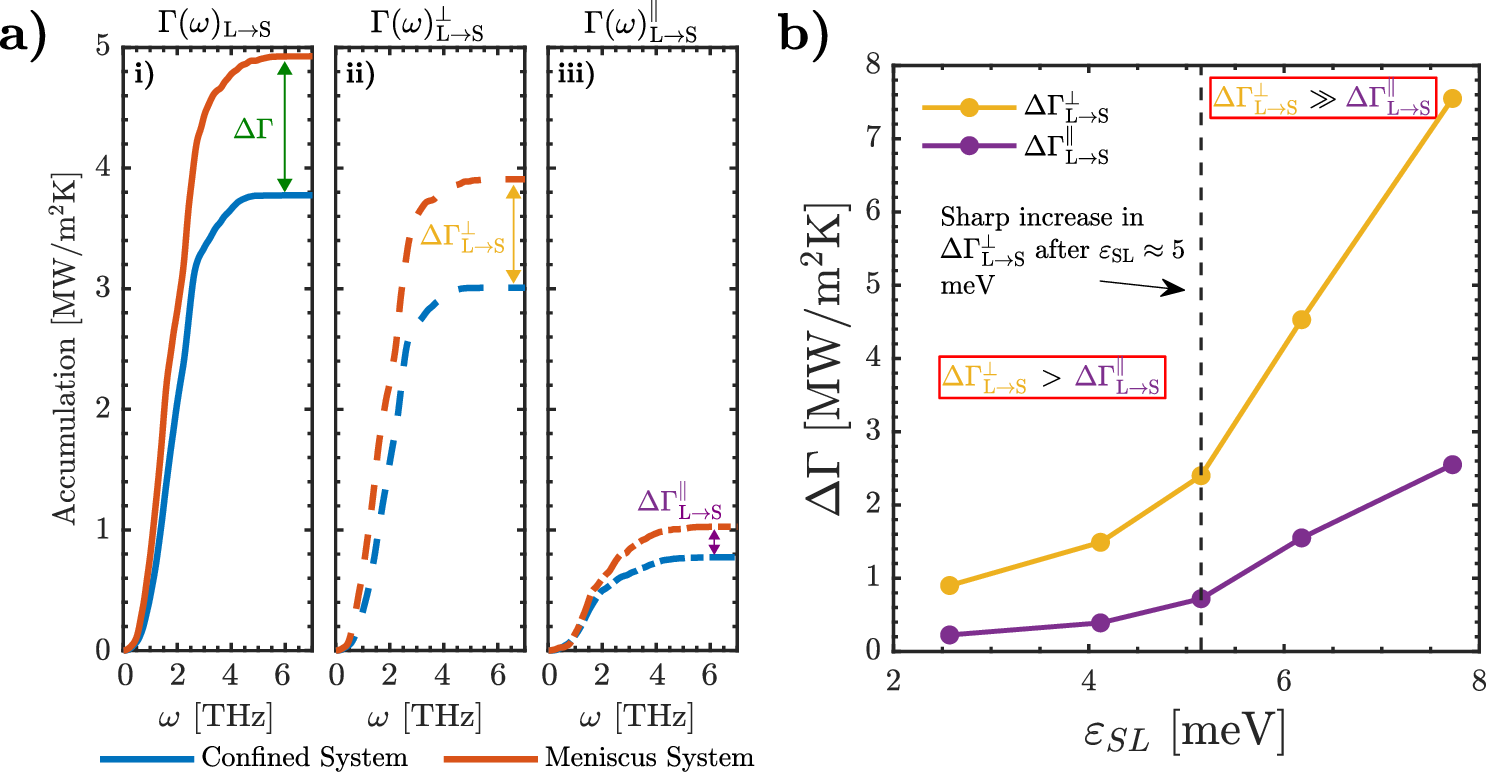}
\caption{(a) Accumulations of the non-normalized spectral decompositions of $\Gamma$ within the interfacial solid for the: (i) total $\Gamma(\omega)_{\rm{L \rightarrow S}}$ and its (ii) out-of-plane ($\Gamma(\omega)_{\rm{L \rightarrow S}}^{\perp}$) and (iii) in-plane ($\Gamma(\omega)_{\rm{L \rightarrow S}}^{\parallel}$) contributions for both the meniscus and confined systems at $\epsilon_{\rm{SL}}=2.5$ meV. The spectral accumulations are obtained via cumulative integration of the original distribution. (b) The relationship between the components of $\Delta \Gamma$ within the interfacial solid ($\Delta \Gamma(\omega)_{\rm{L \rightarrow S}}^{\perp}$, $\Delta \Gamma(\omega)_{\rm{L \rightarrow S}}^{\parallel}$) for varying solid/liquid interaction strength $\epsilon_{\rm{SL}}$.}
\label{fig:PROJ2SDHFnonnormalised}
\end{figure}

Figure \ref{fig:PROJ2SDHFnonnormalised}(a) shows the spectral accumulations of the $\Gamma(\omega)_{\rm{L \rightarrow S}}$ distributions for both the meniscus and confined systems at the lowest value of $\epsilon_{\rm{SL}}=2.5$ meV, accompanied by their further decomposition into their out-of-plane ($\Gamma(\omega)_{\rm{L \rightarrow S}}^{\perp}$) and in-plane ($\Gamma(\omega)_{\rm{L \rightarrow S}}^{\parallel}$) components. The panels thus present the following information:

\begin{enumerate}
    \item The leftmost panel (Figure \ref{fig:PROJ2SDHFnonnormalised}(a)(i)) illustrates the accumulations of the total $\Gamma(\omega)_{\rm{L \rightarrow S}}$ distributions for both systems. The final value of each accumulation represents the actual magnitude of $\Gamma$ of a given system, and the gap between the accumulations of the meniscus and confined systems represents the enhancement in $\Gamma$ due to the meniscus $\Delta \Gamma$, which are identical to the values first shown in Figure \ref{fig:PROJ2Gammameniscusconfined}. 
    
    \item The middle panel (Figure \ref{fig:PROJ2SDHFnonnormalised}(a)(ii)) shows the accumulations of the out-of-plane component of energy transport within the interfacial solid $\Gamma(\omega)_{\rm{L \rightarrow S}}^{\perp}$ for the two systems. Here, the gap between their accumulations represents the difference in the contributions of out-of-plane modes induced by the meniscus, denoted as $\Delta \Gamma _{\rm{IS}} ^{\perp}$ for the interfacial solid.
    
    \item The rightmost panel (Figure \ref{fig:PROJ2SDHFnonnormalised}(a)(iii)) illustrates the accumulations of the in-plane component of heat transfer within the interfacial solid $\Gamma(\omega)_{\rm{L \rightarrow S}}^{\parallel}$ for the meniscus and confined systems. Conversely, the gap between these accumulations represents the difference in the contributions of in-plane modes due to the presence of a meniscus, denoted as $\Delta \Gamma _{\rm{IS}} ^{\parallel}$ for the interfacial solid.
\end{enumerate}

\noindent Analyzing Figures \ref{fig:PROJ2SDHFnonnormalised}(a)(i)-(iii), significant differences can be observed between the spectral accumulations of $\Gamma(\omega)_{\rm{L \rightarrow S}}^{\perp}$ and $\Gamma(\omega)_{\rm{L \rightarrow S}}^{\parallel}$ across the confined and meniscus systems at the lowest value of $\epsilon_{\rm{SL}}=2.5$ meV. Specifically, the meniscus is seen to yield a large increase in $\Gamma(\omega)_{\rm{L \rightarrow S}}^{\perp}$, but a relatively smaller increase in $\Gamma(\omega)_{\rm{L \rightarrow S}}^{\parallel}$.

By quantifying the magnitude of the gaps between the accumulations of $\Gamma(\omega)_{\rm{L \rightarrow S}}^{\perp}$ and $\Gamma(\omega)_{\rm{L \rightarrow S}}^{\parallel}$ for the meniscus and confined systems, $\Delta \Gamma$ can be effectively decomposed within the interfacial solid into its out-of-plane ($\Delta \Gamma(\omega)_{\rm{L \rightarrow S}}^{\perp}$) and in-plane ($\Delta \Gamma(\omega)_{\rm{L \rightarrow S}}^{\parallel}$) components. Figure \ref{fig:PROJ2SDHFnonnormalised}(b) plots these decomposed values within the solid at each value of $\epsilon_{\rm{SL}}$ studied, as derived from their spectral accumulations. From this plot, two observations can be made:

\begin{enumerate}
    \item $\Delta \Gamma(\omega)_{\rm{L \rightarrow S}}^{\perp}$ is significantly larger than $\Delta \Gamma(\omega)_{\rm{L \rightarrow S}}^{\parallel}$ across all magnitudes of $\epsilon_{\rm{SL}}$ considered. Thus, the enhancement in $\Gamma$ due to the presence of a meniscus coincides with the increase in the utilization of out-of-plane modes in the interfacial solid across all values of $\epsilon_{\rm{SL}}$ studied. 
    
    \item  When $\epsilon_{\rm{SL}}<5$ meV, $\Delta \Gamma(\omega)_{\rm{L \rightarrow S}}^{\perp}$ and $\Delta \Gamma(\omega)_{\rm{L \rightarrow S}}^{\parallel}$ both increase monotonously. However, beyond $\epsilon_{\rm{L \rightarrow S}}>5$ meV, both $\Delta \Gamma(\omega)_{\rm{L \rightarrow S}}^{\perp}$ and $\Delta \Gamma(\omega)_{\rm{L \rightarrow S}}^{\parallel}$ rise sharply. This sharp increase in the utilization of the in-plane and out-of-plane modes of the solid coincides with the sharp enhancement in $\Delta \Gamma$ that occurs around $\epsilon_{\rm{SL}} \approx 5$ meV, initially seen in Figure \ref{fig:PROJ2Gammameniscusconfined}(b).

\end{enumerate}

Note that these differences arise from the \textit{transmitted} spectra (i.e.. within the interfacial solid); next, we evaluate the influence of the meniscus on the modes utilized to \textit{receive} energy by the liquid.

\subsubsection{Spectral Accumulations within the Interfacial Liquid}

\begin{figure}[t!]
\centering
\includegraphics[width=\linewidth]{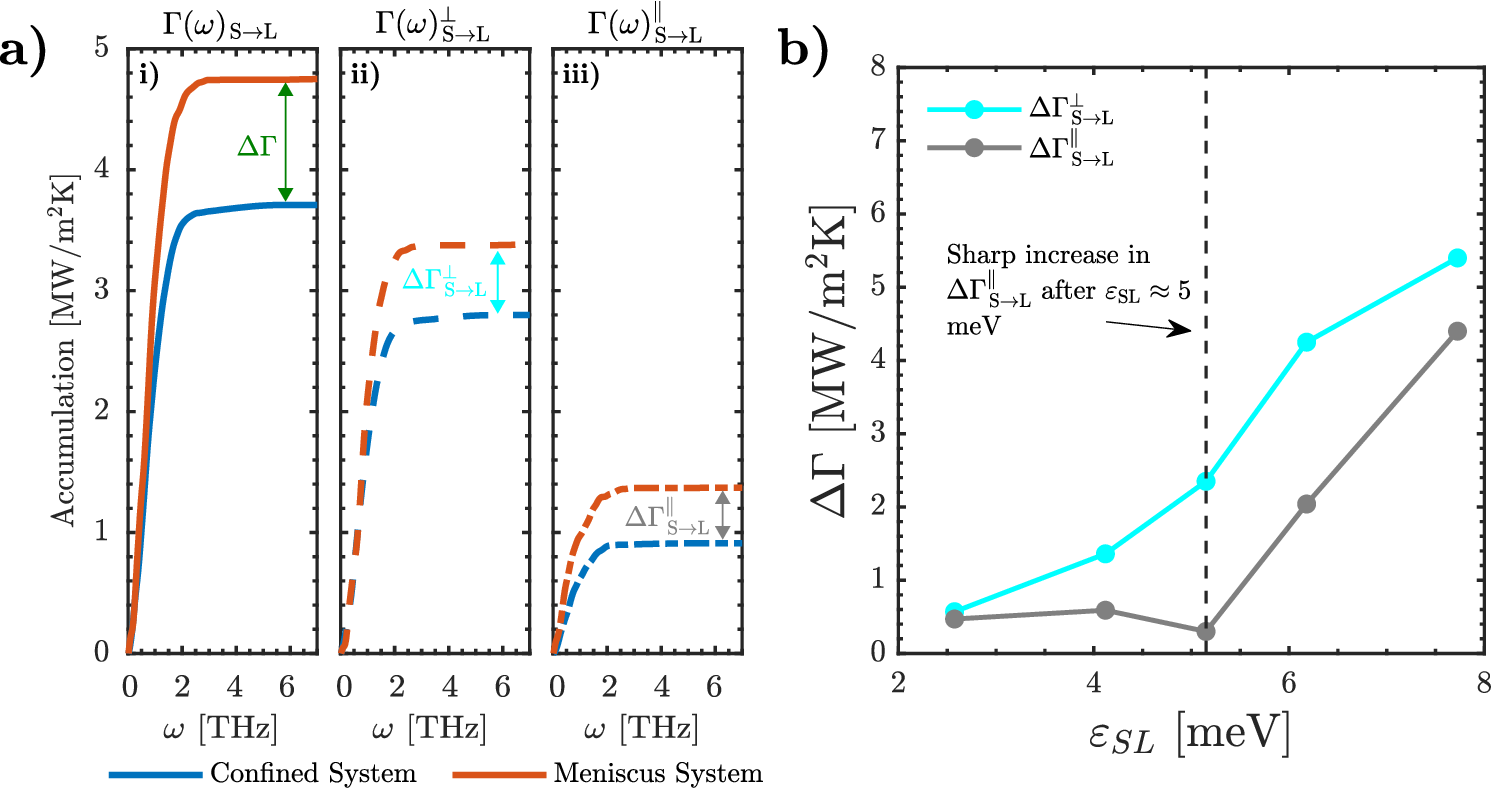}
\caption{(a) Accumulations of the non-normalized spectral decompositions of $\Gamma$ within the interfacial liquid for the: (i) total $\Gamma(\omega)_{\rm{S \rightarrow L}}$ and its (ii) out-of-plane ($\Gamma(\omega)_{\rm{S \rightarrow L}}^{\perp}$) and (iii) in-plane ($\Gamma(\omega)_{\rm{S \rightarrow L}}^{\parallel}$) contributions for both the meniscus and confined systems at $\epsilon_{\rm{SL}}=2.5$ meV. (b) The relationship between the components of $\Delta \Gamma$ within the interfacial liquid ($\Delta \Gamma(\omega)_{\rm{S \rightarrow L}}^{\perp}$, $\Delta \Gamma(\omega)_{\rm{S \rightarrow L}}^{\parallel}$) for varying solid/liquid interaction strength $\epsilon_{\rm{SL}}$.}
\label{fig:liquidaccum}
\end{figure}

Figures \ref{fig:liquidaccum}(a)(i)-(iii) similarly depict the accumulations of the $\Gamma(\omega)_{\rm{S \rightarrow L}}$ distributions of the interfacial liquid for both the meniscus and confined systems at $\epsilon_{\rm{SL}}=2.5$ meV, alongside their out-of-plane ($\Gamma(\omega)_{\rm{S \rightarrow L}}^{\perp}$) and in-plane ($\Gamma(\omega)_{\rm{S \rightarrow L}}^{\parallel}$) components. As was the case with the interfacial solid, the presence of a meniscus gives rise to significant differences in the accumulations of $\Gamma(\omega)_{\rm{S \rightarrow L}}^{\perp}$ and $\Gamma(\omega)_{\rm{S \rightarrow L}}^{\parallel}$ between the confined and meniscus systems at the lowest value of $\epsilon_{\rm{SL}}$; thus, this process is repeated across the entire range of values for $\epsilon_{\rm{SL}}$.

Figure \ref{fig:liquidaccum}(b) similarly depicts the decomposition of $\Delta \Gamma$ into its out-of-plane ($\Delta \Gamma(\omega)_{\rm{S \rightarrow L}}^{\perp}$) and in-plane ($\Delta \Gamma(\omega)_{\rm{S \rightarrow L}}^{\parallel}$) components within the liquid across the entire range of values for $\epsilon_{\rm{SL}}$ studied. From this plot, four observations can be made: 

\begin{enumerate}
    \item When $\epsilon_{\rm{SL}}=2.5$ meV, $\Delta \Gamma(\omega)_{\rm{S \rightarrow L}}^{\perp} \approx \Delta \Gamma(\omega)_{\rm{S \rightarrow L}}^{\parallel}$. Thus, at this magnitude of $\epsilon_{\rm{SL}}$, the enhancement in $\Gamma$ can be equally divided into the rise in the utilization of both out-of-plane and in-plane modes in the interfacial liquid, contrary to the interfacial solid (where the out-of-plane component dominated).

    \item As $\epsilon_{\rm{SL}}$ is increased to 5 meV, $\Delta \Gamma(\omega)_{\rm{S \rightarrow L}}^{\perp}$ rises monotonously, while $\Delta \Gamma(\omega)_{\rm{S \rightarrow L}}^{\parallel}$ remains roughly unchanged (the small dip at 5 meV is within margin of statistical error). Consequently, when $\epsilon_{\rm{SL}}$ is in the range 2.5-5 meV, $\Delta \Gamma(\omega)_{\rm{S \rightarrow L}}^{\perp}$ is notably larger than $\Delta \Gamma(\omega)_{\rm{S \rightarrow L}}^{\parallel}$, and the enhancement in $\Gamma$ yielded by the meniscus coincides with an increase in the utilization of the out-of-plane modes of the liquid. This is in alignment with the interfacial solid in this range of values for $\epsilon_{\rm{SL}}$.

    \item Upon further increasing $\epsilon_{\rm{SL}}$ beyond 5 meV, $\Delta \Gamma(\omega)_{\rm{S \rightarrow L}}^{\parallel}$ rises sharply, while $\Delta \Gamma(\omega)_{\rm{S \rightarrow L}}^{\perp}$ increases to a lesser degree. Consequently, the difference between the two shrinks. As a result, at the highest value of $\epsilon_{\rm{SL}}$, the enhancement in $\Gamma$ coincides with a rise in the utilization of both the in-plane and out-of-plane modes of the interfacial liquid, with the latter being more important.

    \item The sharp increase in the utilization of the in-plane modes of the liquid $\Delta \Gamma(\omega)_{\rm{S \rightarrow L}}^{\parallel}$ around $\epsilon_{\rm{SL}} \approx 5$ meV coincides with the rise in $\Delta \Gamma$ that occurs around the same point, initially observed in Figure \ref{fig:PROJ2Gammameniscusconfined}(b). 
    
    
\end{enumerate}

From this set of spectral analyses, it is evident that the presence of a meniscus significantly impacts the orientation of the utilized modes within both the interfacial solid and liquid, which coincides with the observed enhancement in $\Gamma$. It is also evident that there are differences in spectral transmission between menisci  for different $\epsilon_{\rm{SL}}$. These are discussed in the next section.

\subsection{Connecting Spectral Mechanisms to the Presence and Curvature of the Meniscus}

\begin{figure}[htb!]
    \centering
    \includegraphics[width=\linewidth]{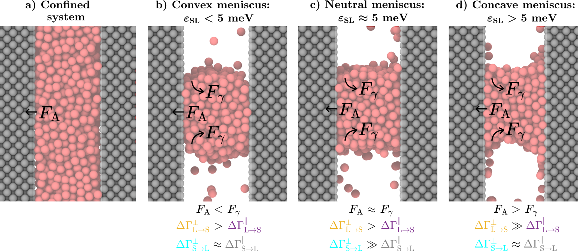}
    \caption{A schematic showing the (a) confined system, demonstrating the adhesive force $F_{\rm{A}}$; (b) convex meniscus for $\epsilon_{\rm{SL}} < 5$ meV, with the adhesive force $F_{\rm{A}}$ and interfacial tension $F_{\gamma}$ illustrated; (c) neutral meniscus for $\epsilon_{\rm{SL}} \approx 5$ meV; and (d) concave meniscus for $\epsilon_{\rm{SL}} > 5$ meV. The relative magnitude of $F_{\rm{A}}$ and $F_{\gamma}$ in each case, as well as the dominant in-plane and out-of-plane mode increases within the solid and liquid, are included below (b)-(d).}
    \label{fig:confinedplusmeniscuscurvatures}
    \end{figure}

The large variation in $\epsilon_{\rm{SL}}$ leads to the formation of menisci of vastly different curvatures, as shown in Figure \ref{fig:confinedplusmeniscuscurvatures}(b)-(d). To understand the relationship between the spectral mechanisms described thus far and the curvature of the meniscus, the interfacial interactions between the three phases at the TPCL of each meniscus must first be examined. Following this, the role of the TPCL in inducing the spectral differences between the confined and meniscus systems will be described. Finally, the origins of the variation in these spectral differences induced by the change in the curvature of the TPCL will be discussed. 

\subsubsection{Interfacial Interactions at the TPCL} 

In the confined system (Figure \ref{fig:confinedplusmeniscuscurvatures}(a)), the liquid experiences an attractive force of adhesion towards the solid, resulting from the solid/liquid interfacial interactions. However, in the meniscus system the liquid/vapor interfacial liquid are pulled by interfacial tension towards the bulk of the meniscus. This liquid/vapor interfacial tension $F_{\gamma}$ is independent of the magnitude of $\epsilon_{\rm{SL}}$, and for the liquid modeled in this work, $F_{\gamma} \approx 17.5$ mN/m.\cite{zhou2011theoretical} Meanwhile, the length-averaged force of adhesion $F_{\rm{A}}$ is a function of $\epsilon_{\rm{SL}}$, and can be estimated from $F_{\gamma}$ and the wetting angle $\theta$ using:

\begin{equation}
\label{eq:workofadh}
    F_{\rm{A}} = (1+\cos\theta)F_{\gamma}
\end{equation}

Using Equation \ref{eq:workofadh} and the wetting angles presented previously in Figure \ref{fig:contactangles}, $F_{\rm{A}}$ can be estimated for each value of $\epsilon_{\rm{SL}}$, as shown in Figure \ref{fig:workofadhesionvssurfacetension}. Thus, the balance between $F_{\rm{A}}$ and $F_{\gamma}$ with varying $\epsilon_{\rm{SL}}$ gives rise to three distinct meniscus curvatures:

\begin{enumerate}
    \item \textbf{Convex Meniscus:} At the lowest value of $\epsilon_{\rm{SL}}$, it can be observed that the meniscus exhibits a pronounced convex shape, with a high curvature of the TPCLs it comprises, as illustrated in Figure \ref{fig:confinedplusmeniscuscurvatures}(b). This is caused by the magnitude of the interfacial tension $F_{\gamma}$ exceeding the magnitude of the solid/liquid adhesive forces $F_{\rm{A}}$ (i.e. $F_{\rm{A}} < F_{\gamma}$), as observed in Figure \ref{fig:workofadhesionvssurfacetension}. Due to these weak adhesion, the TPCL is not ``pinned'' to the surface and can move, leading to a high slip of the interfacial liquid.\cite{Geng2019Slip}

    \begin{figure}[htb!]
    \centering
    \includegraphics[width=0.5\linewidth]{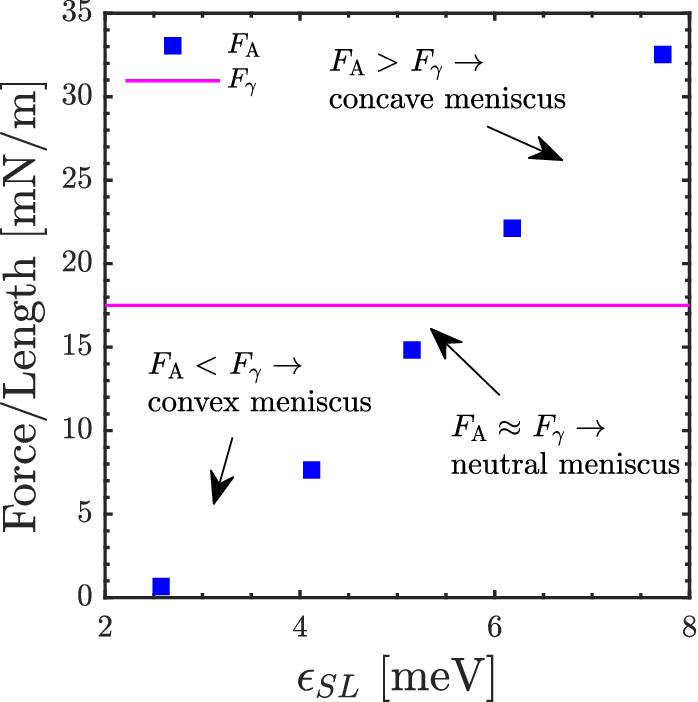}
    \caption{The variation of the length-averaged force of adhesion $F_{\rm{A}}$ with the solid/liquid interaction strength $\epsilon_{\rm{SL}}$ relative to the liquid/vapor interfacial tension $F_{\gamma}$.}
    \label{fig:workofadhesionvssurfacetension}
    \end{figure}

    \item  \textbf{Neutral Meniscus:} As $\epsilon_{\rm{SL}}$ is increased to 5 meV, the meniscus gradually reduces the curvature of its TPCLs, taking a neutral shape around $\epsilon_{\rm{SL}} \approx$ 5 meV, as shown in Figure \ref{fig:confinedplusmeniscuscurvatures}(c). In this case, the magnitude of the interfacial tension $F_{\gamma}$ is roughly equal to that of the adhesive forces $F_{\rm{A}}$ (i.e. $F_{\rm{A}} \approx F_{\gamma}$), as seen in Figure \ref{fig:workofadhesionvssurfacetension}. As a result, the interfacial slip reduces, but is still non-negligible.\cite{Mandrolko2024}

    \item \textbf{Concave Meniscus:} As $\epsilon_{\rm{SL}}$ is increased beyond 5 meV, the meniscus begins taking a concave shape, now increasing its curvature once again, as observed in Figure \ref{fig:confinedplusmeniscuscurvatures}(d). This is caused by the magnitude of $F_{\rm{A}}$ exceeding that of $F_{\gamma}$ (i.e. $F_{\rm{A}} > F_{\gamma}$), as shown in Figure \ref{fig:workofadhesionvssurfacetension}. Consequently, the TPCL becomes pinned, and the interfacial slip becomes negligible.\cite{Mandrolko2024}

\end{enumerate}

\subsubsection{Origins of Spectral Differences between Confined and Meniscus Systems}

In the solid, Figure \ref{fig:PROJ2SDHFnonnormalised}(b) showed that the increase in $\epsilon_{\rm{SL}}$ to 5 meV, i.e. the transition from a convex meniscus (Figure \ref{fig:confinedplusmeniscuscurvatures}(b)) to a neutral meniscus (Figure \ref{fig:confinedplusmeniscuscurvatures}(c)), coincides with a rise in the utilization of the out-of-plane modes of the solid ($\Delta \Gamma(\omega)_{\rm{L \rightarrow S}}^{\perp}$), accompanied by a smaller increase in the utilization of its in-plane modes ($\Delta \Gamma(\omega)_{\rm{L \rightarrow S}}^{\parallel}$). Further increases in $\epsilon_{\rm{SL}}$ beyond 5 meV, i.e. the transition from a neutral meniscus (Figure \ref{fig:confinedplusmeniscuscurvatures}(c)) to a concave one (Figure \ref{fig:confinedplusmeniscuscurvatures}(d)), coincide with sharp rises in the utilization of the out-of-plane ($\Delta \Gamma(\omega)_{\rm{L \rightarrow S}}^{\perp}$) and in-plane ($\Delta \Gamma(\omega)_{\rm{L \rightarrow S}}^{\parallel}$) modes of the solid.

In the liquid, Figure \ref{fig:liquidaccum}(b) showed that the transition from a convex meniscus to a neutral one corresponds to a rise in the utilization of the out-of-plane modes of the liquid ($\Delta \Gamma(\omega)_{\rm{S \rightarrow L}}^{\perp}$), and an insignificant change in the utilization of its in-plane modes ($\Delta \Gamma(\omega)_{\rm{S \rightarrow L}}^{\parallel}$). Beyond $\epsilon_{\rm{SL}} \approx 5$ meV, the utilization of the out-of-plane modes in the liquid continues to increase monotonously, while the utilization of its in-plane modes increases sharply. These enhancements between the confined and meniscus systems, as well as the dependence of these enhancements on the wettability of the surface, can be explained in the following way:

\begin{enumerate}
    \item \textbf{Solid:}
    \begin{enumerate}
        \item Increase in out-of-plane utilization ($\Delta \Gamma(\omega)_{\rm{L \rightarrow S}}^{\perp}$): It has been demonstrated that, at a solid/vapor interface, the out-of-plane modes within the solid are less hindered when compared to a solid/liquid interface, and are thus utilized to a larger degree.\cite{Giri2016ImplicationsStudy} This is consistent with our observations where the utilization of these out-of-plane modes is increased for the meniscus systems where a vapor region is present, when compared to the confined system which is fully liquid. 

        \item Variation of out-of-plane utilization with meniscus shape: With increasing $\epsilon_{\rm{SL}}$, the reduction in the curvature of the meniscus exposes the solid at the TPCL to a larger number of vapor atoms, and a correspondingly smaller number of liquid atoms. Thus, the mechanism in (a) above is drastically magnified upon the transition from a neutral meniscus to a concave one, due to the significant reduction in the number of liquid atoms adjacent to the solid atoms at the TPCL. This likely induces the sharp rise in the utilization of the out-of-plane modes beyond $\epsilon_{\rm{SL}} \approx 5$ meV, previously observed in Figure \ref{fig:PROJ2SDHFnonnormalised}(b).

        \item Increase in in-plane utilization ($\Delta \Gamma(\omega)_{\rm{L \rightarrow S}}^{\parallel}$): The interplay between the adhesive forces $F_{\rm{A}}$ and the interfacial tension $F_{\gamma}$ localized on the solid atoms adjacent to the TPCL enables the solid to utilize its in-plane modes more effectively. 

        \item Variation of in-plane utilization with meniscus shape: With increasing $\epsilon_{\rm{SL}}$, the resulting interfacial forces acting on the solid atoms of the TPCL increase steadily in magnitude,\cite{Fan2020MicroscopicLine} yielding the steady increase in the utilization of the in-plane modes of the solid observed in Figure \ref{fig:PROJ2SDHFnonnormalised}(b) for $\epsilon_{\rm{SL}} < 5$ meV.

    \end{enumerate}


    \item \textbf{Liquid:} 

    \begin{enumerate}
        \item Increase in out-of-plane ($\Delta \Gamma(\omega)_{\rm{L \rightarrow S}}^{\perp}$) and in-plane ($\Delta \Gamma(\omega)_{\rm{L \rightarrow S}}^{\parallel}$) utilization: Similar to the solid, the interplay between the adhesive forces $F_{\rm{A}}$ and the interfacial tension $F_{\gamma}$ localized on the liquid atoms in the TPCL acts as an additional dissipation mechanism for all received modes, facilitating their transmission from the solid into the liquid.

        \item Variation of out-of-plane utilization with meniscus shape: With increasing $\epsilon_{\rm{SL}}$, this mode dissipation is amplified, facilitating the transfer of the out-of-plane modes of the solid into out-of-plane modes within the liquid. This explains the rise in the utilization of the out-of-plane modes of the liquid observed in Figure \ref{fig:liquidaccum}(b).
    \end{enumerate}

\end{enumerate}

\subsubsection{Transition from Convex to Neutral Meniscus}

Figure \ref{fig:liquidaccum}(b) showed that the liquid experiences virtually no increase in the utilization of its in-plane modes ($\Delta \Gamma(\omega)_{\rm{S \rightarrow L}}^{\parallel}$) when $\epsilon_{\rm{SL}}$ is increased to 5 meV, i.e. upon the transition from a convex meniscus (Figure \ref{fig:confinedplusmeniscuscurvatures}(b)) to a neutral meniscus (Figure \ref{fig:confinedplusmeniscuscurvatures}(c)). This occurs because, despite the increase in the magnitude of $F_{\rm{A}}$ relative to $F_{\gamma}$, the interfacial slip that the liquid experiences is still significant,\cite{Mandrolko2024,Geng2019Slip} and its in-plane modes cannot be effectively utilized.\cite{Giri2016ImplicationsStudy,Saaskilahti2016SpectralInterfaces}

\subsubsection{Transition from Neutral to Concave Meniscus}

With further increases in $\epsilon_{\rm{SL}}$, the curvature of the meniscus transition from a neutral meniscus (Figure \ref{fig:confinedplusmeniscuscurvatures}(c)) to a concave meniscus (Figure \ref{fig:confinedplusmeniscuscurvatures}(d)). Figure \ref{fig:PROJ2SDHFnonnormalised}(b) showed that this coincides with a sharp rise in the utilization of the in-plane modes of the solid ($\Delta \Gamma(\omega)_{\rm{L \rightarrow S}}^{\parallel}$), while Figure \ref{fig:liquidaccum}(b) similarly showed a sharp increase in the utilization of the in-plane modes of the liquid ($\Delta \Gamma(\omega)_{\rm{S \rightarrow L}}^{\parallel}$). As discussed prior, the large increase the magnitude of the adhesion force $F_{\rm{A}}$ relative to the interfacial tension $F_{\gamma}$ leads to the pinning of the TPCL, drastically reducing the slip of the interfacial liquid.\cite{Mandrolko2024} This significantly facilitates the utilization of the in-plane modes of the solid and liquid, consistent with prior observations where the transition from phobicity to philicity led to a rise in the utilization of in-plane modes.\cite{Ramos-Alvarado2017SpectralInterfaces,Gonzalez-Valle2019SpectralInterfaces}

\section{Conclusions} 
The impact of a meniscus on the area-normalized interfacial thermal conductance ($\Gamma$) was studied at an LJ solid/liquid interface for various magnitudes of the solid/liquid interaction strength $\epsilon_{\rm{SL}}$. Across all values of $\epsilon_{\rm{SL}}$, the presence of a meniscus was found to yield a significant enhancement in $\Gamma$ ($\Delta \Gamma$). This enhancement was found to rise monotonously until $\epsilon_{\rm{SL}} \approx 5$ meV, beyond which it increases sharply. To understand why the meniscus yields an enhancement in $\Gamma$, and why the magnitude of this enhancement rises sharply beyond $\epsilon_{\rm{SL}} \approx 5$ meV, the spectral decomposition of heat flux formalism was used to probe the interfacial solid and liquid, respectively. 

The meniscus was shown to have negligible influence on the frequencies of the utilized vibrations in the interfacial solid or liquid. Instead, it was seen to influence the preferred orientation of the utilized vibrations. The monotonous enhancement in $\Delta \Gamma$ until $\epsilon_{\rm{SL}} \approx 5$ meV coincided with an increase in the utilization of the out-of-plane vibrations engaged in heat transfer within both the interfacial solid and liquid. This was found to be caused by the interplay between adhesive forces and interfacial tension at the TPCL, which facilitates the utilization of out-of-plane modes in both media. 

The sharp enhancement in $\Delta \Gamma$ beyond $\epsilon_{\rm{SL}} \approx 5$ meV was found to be related to the sharp increases in the utilization of the in-plane vibrations within both the interfacial solid and liquid, accompanied by a sharp rise in the utilization of the out-of-plane modes of the solid. This was found to be related to the drastic reduction in the slip of the interfacial liquid upon the transition from a neutral meniscus to a concave one. These results elucidate heat transfer processes at the TPCL, which are particularly relevant to state-of-the-art two-phase evaporators for integrated circuit cooling, among other applications.

\begin{acknowledgement}
This research is supported by ANR project ``PROMENADE'' No. ANR-23-CE50-0008. Molecular simulations were conducted using HPC resources from GENCI-TGCC and GENCI- IDRIS (No. A0150913052), as well as resources provided by the EXPLOR Center hosted by the University of Lorraine. S.P.\ thanks the Leverhulme Trust for the support provided through the Early Career Fellowship ECF-2021-137.

\end{acknowledgement}



\bibliography{references.bib,books.bib}

\clearpage

\begin{appendices}
\renewcommand{\thefigure}{A\arabic{figure}}

\setcounter{figure}{0}

\begin{figure}[htb!]
\centering
\includegraphics[width=\linewidth]{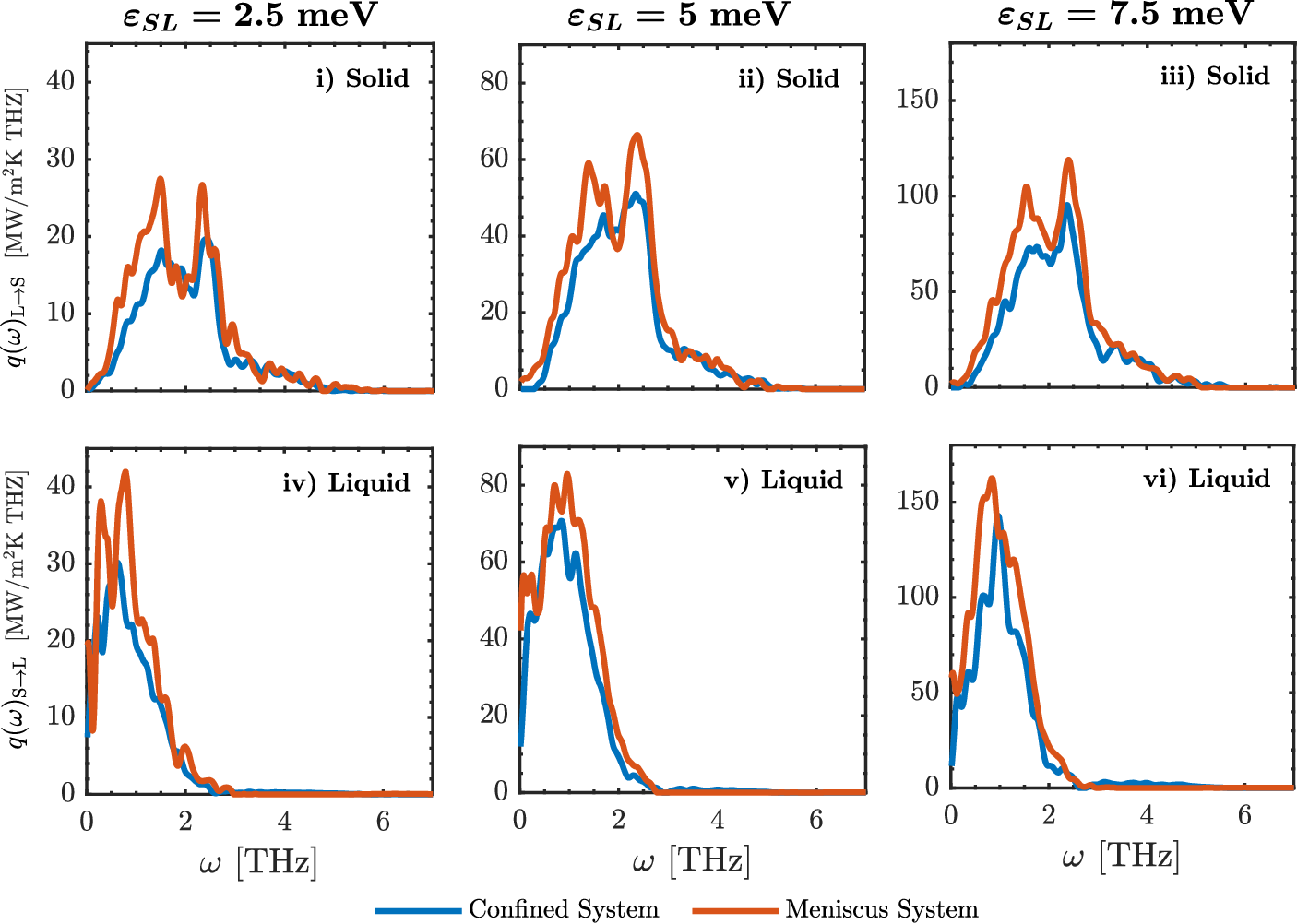}
\caption{Spectral decompositions of heat flux within the interfacial solid ($q(\omega)_{\rm{L \rightarrow S}}$) and liquid ($q(\omega)_{\rm{S \rightarrow L}}$) for the confined and meniscus systems at select values of the solid/liquid interaction strength $\epsilon_{\rm{SL}}$.}
\label{fig:appendixnonnormalised}
\end{figure}

\begin{figure}[htb!]
\centering
\includegraphics[width=\linewidth]{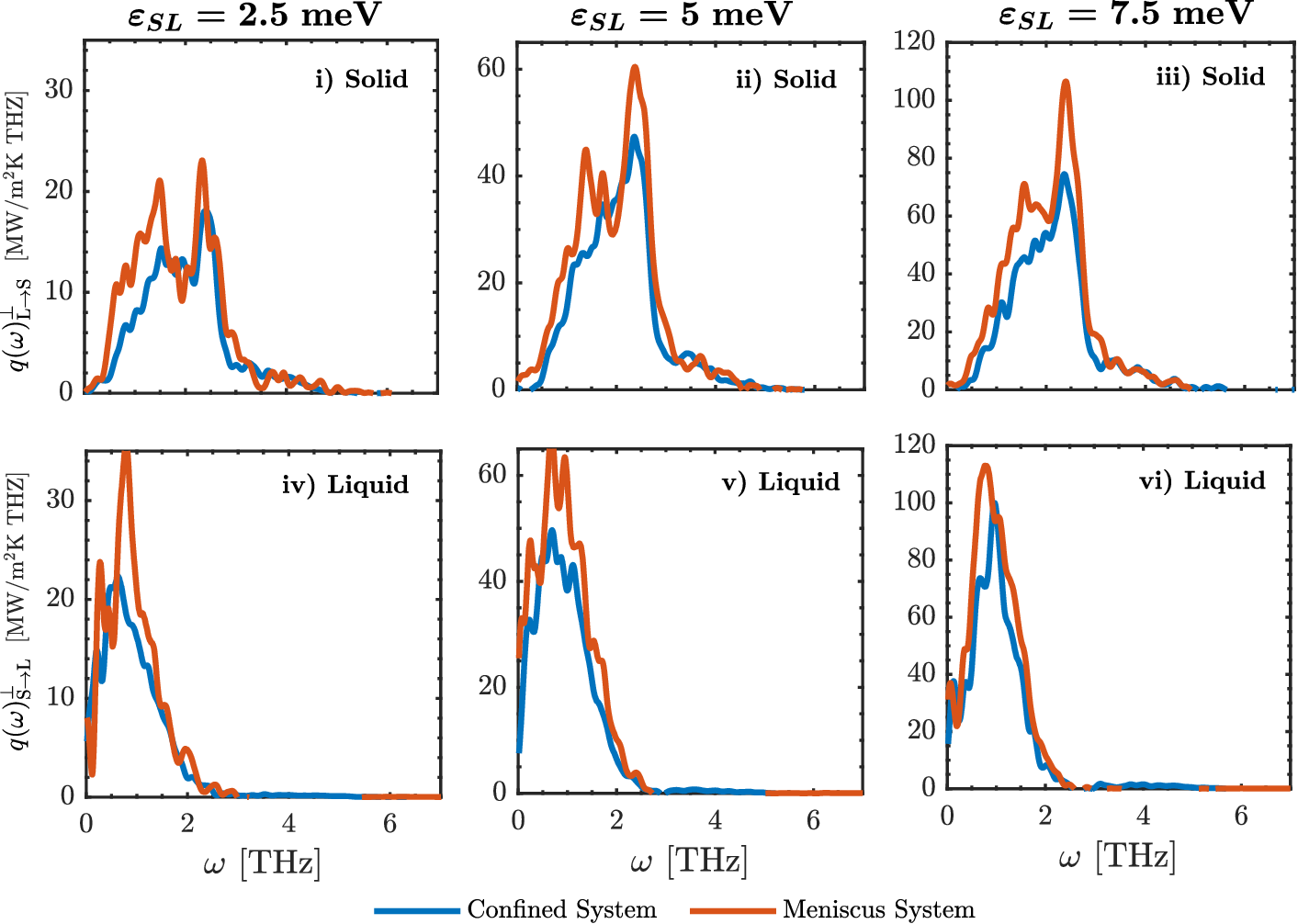}
\caption{Out-of-plane spectral decompositions of heat flux within the interfacial solid ($q(\omega)_{\rm{L \rightarrow S}}^{\perp}$) and liquid ($q(\omega)_{\rm{S \rightarrow L}}^{\perp}$) for the confined and meniscus systems at select values of the solid/liquid interaction strength $\epsilon_{\rm{SL}}$.}
\label{fig:appendixperp}
\end{figure}

\begin{figure}[htb!]
\centering
\includegraphics[width=\linewidth]{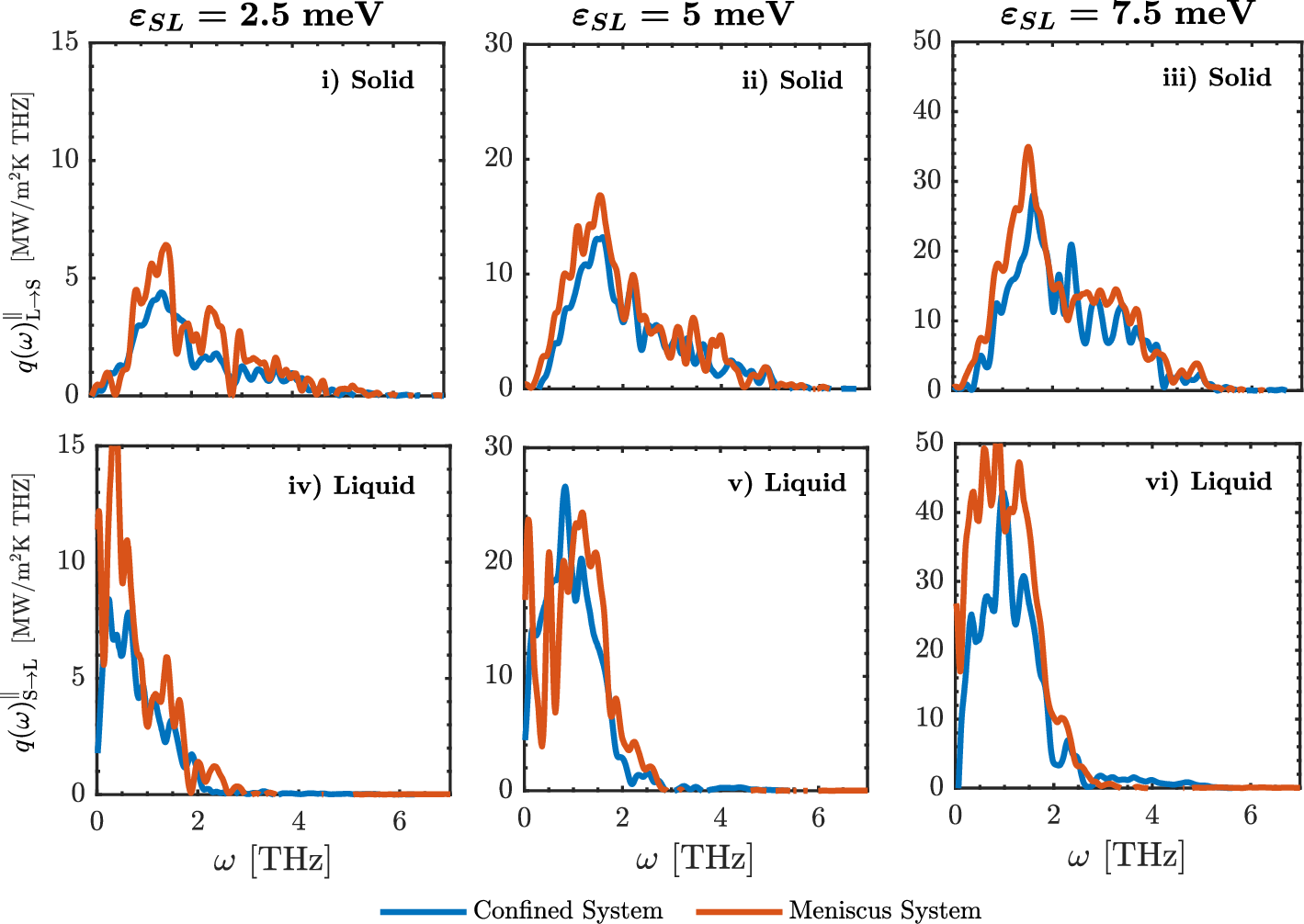}
\caption{In-plane spectral decompositions of heat flux within the interfacial solid ($q(\omega)_{\rm{L \rightarrow S}}^{\parallel}$) and liquid ($q(\omega)_{\rm{S \rightarrow L}}^{\parallel}$) for the confined and meniscus systems at select values of the solid/liquid interaction strength $\epsilon_{\rm{SL}}$.}
\label{fig:appendixparallel}
\end{figure}

\begin{figure}[htb!]
\centering
\includegraphics[width=\linewidth]{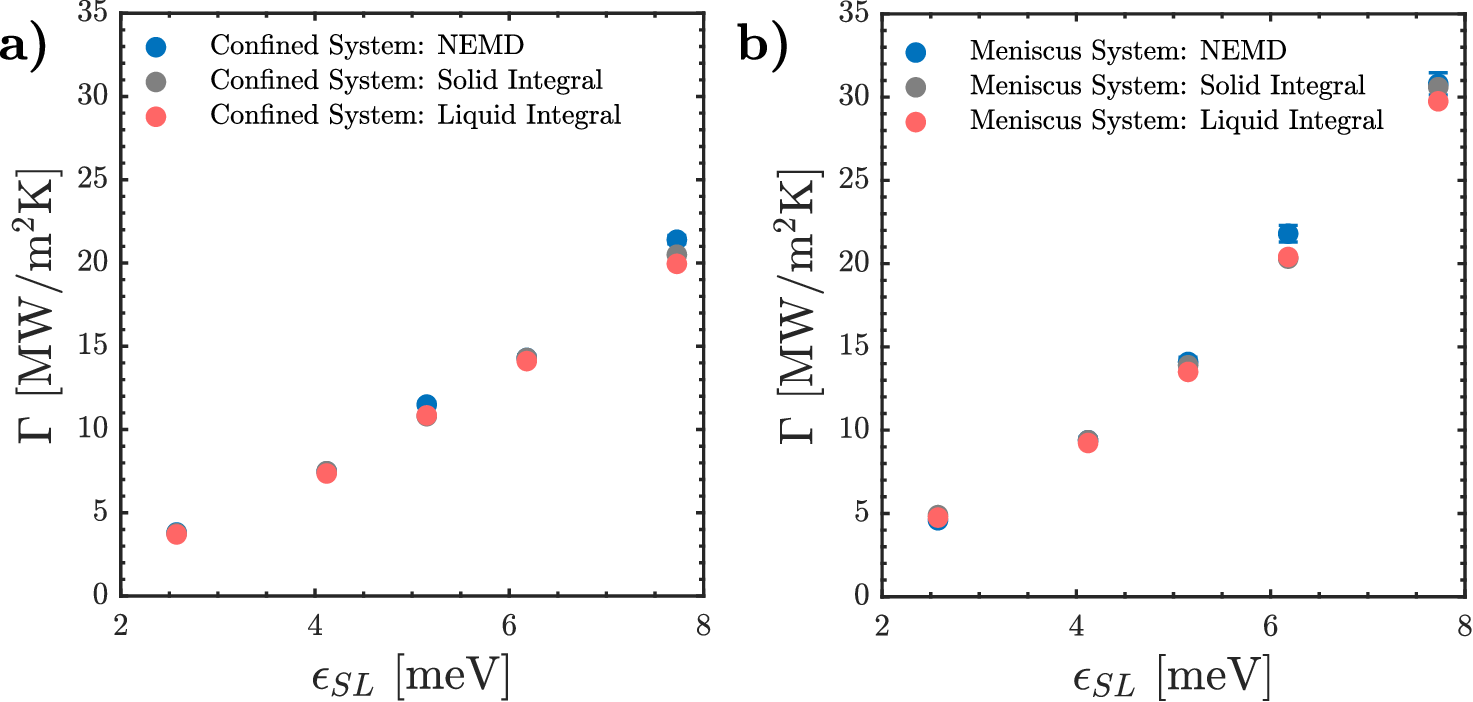}
\caption{(a) Validating the spectral integrals within the interfacial solid and liquid for the confined system across all values of $\epsilon_{\rm{SL}}$. (b) Validating the spectral integrals within the interfacial solid and liquid for the meniscus system across all values of $\epsilon_{\rm{SL}}$.}
\label{fig:appendixvalidation}
\end{figure}

\end{appendices}

\end{document}